\documentclass[preprint,epsf]{aastex}
\def\deg{\ifmmode^\circ\else$^\circ$\fi}

\def\kms{\ifmmode{\rm km\,s^{-1}}\else{km\,s$^{-1}$}\fi}

\slugcomment{Accepted for publication in The Astronomical Journal}
\shortauthors{Balcells et. al.}
\shorttitle{NGC 3656 HI}

\begin{document}

\title{HI in the shell elliptical galaxy NGC 3656\altaffilmark{1}}

\author{Marc Balcells\altaffilmark{2}, 
J. H. van Gorkom\altaffilmark{3},
Renzo Sancisi\altaffilmark{4,5}, and
Carlos del Burgo\altaffilmark{2,6}}

\altaffiltext{1}{Based on observations made with the William Herschel Telescope 
operated on the island of La Palma by the Isaac Newton Group of Telescopes 
in the Spanish Observatorio del Roque de los Muchachos of the Instituto de 
Astrof\'\i sica de Canarias.}
\altaffiltext{2}{Instituto de Astrofisica de Canarias, 
C/ V\'\i a L\'actea, 38200 La Laguna, Canary Islands, Spain}
\altaffiltext{3}{Department of Astronomy, Columbia University, 
	550 W. 120th Street, New York, NY  10027, USA}
\altaffiltext{4}{Osservatorio Astronomico di Bologna, Via Ranzani 1, 
I-40127, Italy}
\altaffiltext{5}{Kapteyn Astronomical Institute, University of Groningen,
	Postbus 800, 9700 AV Groningen, The Netherlands}
\altaffiltext{6}{Max Planck Institut f\"ur Astronomie, K\"onigstuhl 
17, D-69117 Heidelberg, Germany}

\begin{abstract}
Very Large Array\footnote{The VLA of the National Radio Astronomy
Observatory is operated by Associated Universities, Inc.  under
cooperative agreement with the National Science Foundation.} 
(VLA) neutral hydrogen observations of the shell elliptical NGC~3656 reveal
an edge-on, warped minor axis gaseous disk ($M_{HI} \sim$2$\times
10^9$ M$_{\odot}$) extending 7 kpc.  HI is also found outside the
optical image, on two complexes to the North-East and North-West 
that seem to trace an outer broken HI 
disk or ring, or possibly one or two tidal tails.  
These complexes link with the outer edges of the inner disk,
and appear displaced with respect to the two optical tails in the
galaxy.  The disk kinematics is strongly lopsided, suggesting recent
or ongoing accretion.

Integral-field optical fiber spectroscopy at the region of the bright
southern shell of NGC~3656 has provided a determination of the
stellar velocities of the shell.  The shell, at 9 kpc from
the center, has traces of HI with velocities bracketing the stellar
velocities, providing evidence for a dynamical association of HI and
stars at the shell.  Within the errors the stars have systemic
velocity, suggesting a possible phase wrapping origin for the shell.

We probed a region of 40'$\times$40' (480 kpc$\times$480 kpc) $\times$
1160 km/s down to an HI mass sensitivity (6 $\sigma$) of $3 \times
10^7 M_{\odot}$ and detect five dwarf galaxies with HI masses ranging
from $2\times 10^8 M_{\odot}$ to $2\times 10^9 M_{\odot}$ all within 180 kpc from
NGC 3656 and all within the velocity range (450 \kms) of the HI of NGC
3656.  The dwarfs were previously catalogued but none had a known
redshift.  For the NGC 3656 group to be bound requires a total mass of
$3-7.4 \times 10^{12} M_{\odot}$, yielding a mass to light ratio from 125 to 300.

The overall HI picture presented by NGC~3656 supports the hypothesis of
a disk-disk merger origin, or possibly an ongoing process of multiple
merger with nearby dwarfs.
\end{abstract}

\keywords{Galaxies: elliptical and lenticular --- galaxies: kinematics and
dynamics --- galaxies: interactions --- galaxies: peculiar --- galaxies:
individual (NGC~3656, UGC 6400, UGC 6422, MCG +09-19-052, MCG +09-19-056, MCG +09-19-059, MAPS-NGP O\_130\_0507168) }

\section{Introduction}
\label{Sec:Introduction}

\begin{figure}
%\epsscale{0.8}
\plotone{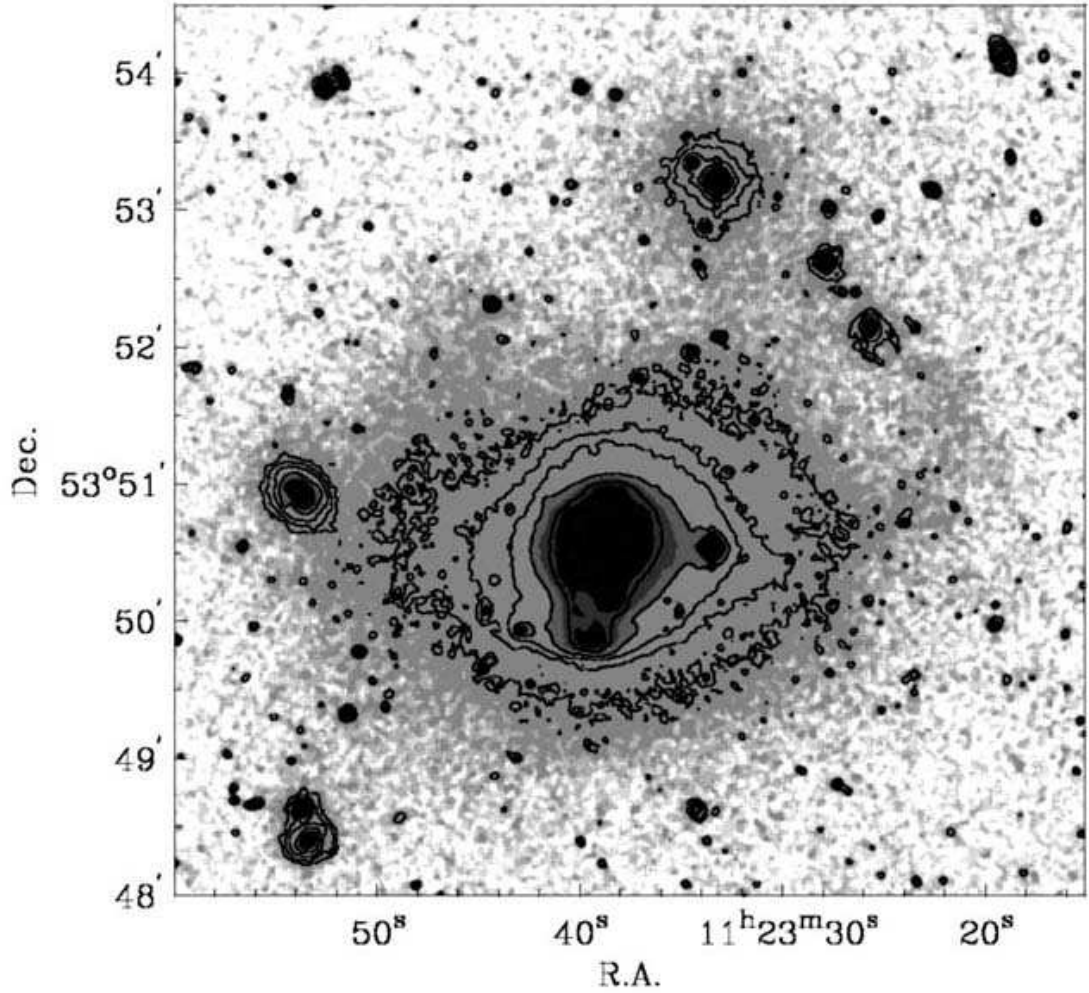}
\caption{ \label{Fig:DeepR} R-band image of NGC~3656 (from Fig.~1 of 
B97).  Lowest contour is 26.0 mag/arcsec$^2$,
contour spacing is 0.75 mag.  
Gray levels, drawn after applying a 5$\times$5 pixel median
filter,  correspond to 
$sky - 3\sigma$ ({\it white}), $sky - 2\sigma$, $sky - \sigma$,
$sky + \sigma, sky + 2\sigma, sky + 3\sigma$, to show the faintest 
extended optical structures.  Flatfielding of the image is better than 
0.2\%.
Objects at $(\alpha, \delta)$ = 
(11:23:33, 53:50:33), (11:23:33.1, 53:53:13) and
(11:23:27.7,53:52:38) are stars. Scale is 0.2 kpc\,arcsec$^{-1}$.  }
\end{figure}

NGC 3656 (Arp~155) is a peculiar elliptical galaxy with a nearly
spherical body and a dark band running North-South (see
Figure~\ref{Fig:DeepR}).  A light clump can be seen 45 arcsec 
South from the center, which is bound to the South by a prominent shell. 
A number of other, nearly circular shells are seen around the galaxy. 
The core rotation axis is orthogonal to that of the main body
(Balcells \& Stanford 1990, hereafter BS90).  Two faint optical tails
exist around the galaxy (Balcells 1997, hereafter B97).  Color maps of
the galaxy (B97) show that the shell to the South is
distinctly blue ($B-R \approx 0.95$) in relation to the surroundings
($B-R \approx 1.4$), pointing to a young age for its stars.  
Star formation ($\sim 0.1$ M$_{\odot}$\,yr$^{-1}$) is ongoing at the dust lane, as evidenced by
extended emission in H$\alpha$ (B97) and radio continuum
(M\"ollenhoff, Hummel \& Bender 1992).

Observations of the 21-cm hydrogen emission line obtained with the
Westerbork Synthesis Radio Telescope (WSRT) show that the galaxy
contains about $10^9$ M$_\odot$ of neutral hydrogen (Balcells \&
Sancisi 1996, hereafter BS96).  The HI distribution peaks on the dust
lane and extends in the North-South direction.  The gas on the dust
lane shows rapidly rising rotation with velocities reaching up to
about 225 \kms, supporting the hypothesis of a rotating disk seen
almost edge-on.  In addition to this HI component, there is a central
concentration of molecular gas of similar mass (BS96).

NGC~3656 is useful for a study of the origin of gaseous disks in
ellipticals because of the large number of peculiarities normally
ascribed to interactions that coincide in this galaxy: shells, two
tails, peculiar kinematics, peculiar colors, star formation.  These
can provide constraints on formation scenarios for the gaseous disk. 
In addition, NGC~3656 is interesting for a study of the HI-shell relation,
because of the apparently young dynamical age of the system and
because the southern shell is
significantly brighter (1.5 mag/arcsec$^2$) than the galaxy background
at that distance from the center.

We present here VLA aperture synthesis HI observations and
integral-field spectroscopy (IFS) using fiber optics on NGC~3656.  HI
measurements are used to describe the gas distribution and velocities
with higher resolution and sensitivity than achieved with the WSRT,
and to look for the existence of diffuse gas around the galaxy.  The
IFS of the shell to the south have provided an optical
velocity determination for stars associated with the shell.

The optical IFS is presented in \S~\ref{Sec:2Dspectroscopy}. 
Section~\ref{Sec:HIobservations} describes the HI observations. 
Results are presented in \S~\ref{Sec:Results}, for the inner HI disk
(\S~\ref{Sec:InnerDisk}), for the extended HI
(\S~\ref{Sec:ExtendedEmission}) and for the shell kinematics
(\S~\ref{Sec:ShellKinematics}).  We discuss the HI results in
\S~\ref{Sec:Discussion}, the connection between HI and the shell in
\S~\ref{Sec:ShellDynamics}, and possible formation scenarios for
NGC~3656 in \S~\ref{Sec:Formation}.  A Hubble constant of 75 \kms
Mpc$^{-1}$ and a distance of 40 Mpc are assumed throughout, giving a
scale of 0.2 kpc arcsec$^{-1}$.  All coordinates refer to the J2000.0
equinox.

\section{Integral-field optical fiber spectroscopy}
\label{Sec:2Dspectroscopy}

We observed the southern shell of NGC 3656 with the William Herschel
telescope, on 1998 March 31, using the INTEGRAL fiber bundle system
which feeds the WYFFOS Nasmyth spectrograph.  For a description of
INTEGRAL, see Arribas et al.\ (1998).  We used the SB3 bundle, which
contains 135 fibers, each 600 $\mu$m in diameter (2.7 arcsec on the
sky).  Of these, 115 densely cover an area of
34$^{\prime\prime}\times$30$^{\prime\prime}$ on the sky.  The
distribution of these fibers is hexagonal with a mean distance between
them of 3 arcsec.  The remaining 20 fibers are set in a ring 45 arcsec
in diameter, which is generally used for a sky measurement.  We used
grating R1200R, which gives a mean dispersion of 1.4 \AA\,pixel$^{-1}$
in the spectral range 5640-7070 \AA\ (covering the NaI doublet at 5890
\AA, the 6563 \AA\ H$\alpha$ line, and the [SII]6716 and 6731 \AA\
lines).  The optical fiber bundle was positioned at ($\alpha$,
$\delta$) = (11$^{\rm h}$23$^{\rm m}$38$^{\rm s}$.6, 53$^{\rm
o}$49$^{\prime}$57$^{\prime\prime}$.6), $\sim$35~arcsec 
south of the galaxy center, and covers from the southern end of the
dust lane to beyond the southern end of the shell.  The region covered
by the bundle is indicated with a thick rectangle in
Figure~\ref{Fig:TotHI7}.

\begin{figure}
\epsscale{0.7}
\plotone{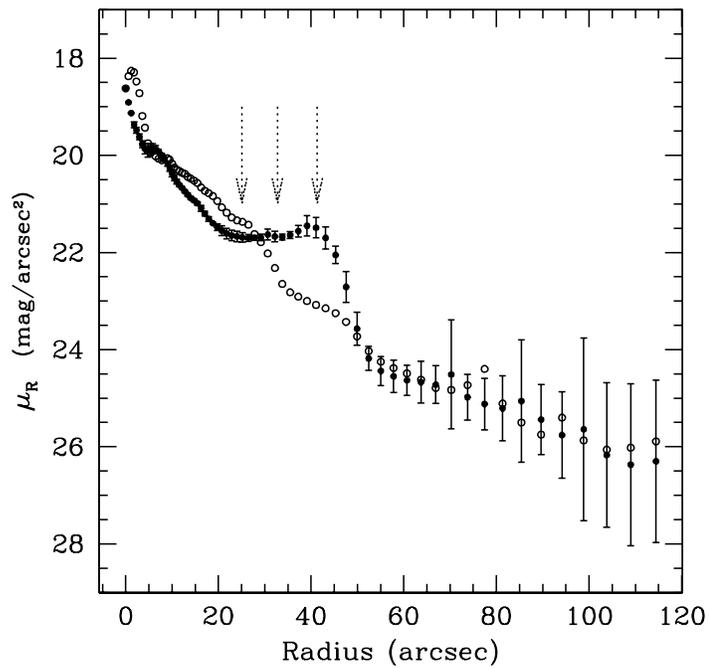}
\caption{ \label{Fig:RbandProfiles} R-band surface brightness profiles 
of NGC~3656 on 25$^\circ$-wide wedge-shaped apertures centered on the 
galaxy nucleus, along PA=170$^\circ$ (filled symbols, with error bars) and 
along PA=10$^\circ$ (open symbols, error bars omitted).  Vertical arrows 
indicate the central positions of the apertures where optical velocities 
from 2D fiber spectroscopy are measured.}
\end{figure}

We followed the data reduction procedure described in del Burgo (2000):
bias subtraction, aperture definition and trace, stray light subtraction,
extraction of the apertures, wavelength calibration, throughput correction,
sky subtraction and cosmic-ray rejection.  The throughput correction was
done from sky and dome flat exposures.  Sky subtraction was accurate to 
better than 5\% rms.  

The individual spectra show the NaI absorption doublet and traces of
H$\alpha$, NII and SII emission lines.  The weakness of these lines
indicates a very slow rate of star formation activity in the
shell at present.  However, H$\alpha$ absorption indicates that
star formation must have been important in the past $\sim 10^9$ yr. 
Overall, the distribution of the H$\alpha$ emission matches the one
displayed in the H$\alpha$ narrow-band image in Figure~2b of B97.

We obtained stellar velocities via cross-correlation, using the spectrum of
a G2V star as a template.  We opted not to use the H$\alpha$ line because,
in many fibers, this is partially filled with H$\alpha$ emission.  Instead,
we used the region around 5890 \AA\ which includes the NaI doublet.  We
cross-correlated each fiber spectrum, and excluded those where no
convergence was found.  We then coadded groups of spectra that showed
similar velocities, and cross-correlated these with the template to verify
that the results do not depend on the signal-to-noise ratio of the spectra. 
Because of the intrinsic low signal of each spectrum, the individual fiber
velocities have errors around 100 \kms.  Coadded spectra yield velocities
with errors around 60 \kms.  The errors are probably conservative as
cross-correlation is known to overestimate velocity errors (BS90;
Jedrzejewski \& Schechter 1988).  This yielded three velocity values
centered at positions 25, 33 and 41 arcsec South from the center.  These
synthesized apertures are indicated with rectangles in
Figure~\ref{Fig:TotHI7}.

To show the three optical velocity apertures with respect to the
southern shell, we plot $R$-band surface brightness profiles along
wedge-shaped apertures along PA=170$^{\circ}$ and PA=10$^{\circ}$,
centered on the galaxy nucleus (Figure~\ref{Fig:RbandProfiles}).  The
vertical arrows locate the center positions of the three fiber velocity data
apertures.  We take the North profile as an estimate of the underlying
galaxian surface brightness distribution, and use this to measure the
surface brightness contrast at the South shell.  The latter, at $R=45$
arcsec, has $\mu_R=21.5$ mag/arcsec$^{-2}$, 1.5 mag arcsec$^{-2}$
brighter than the background.  The outermost velocity point is
centered on the South shell, thus the measurement traces the
shell velocity with little background contamination.  At the locations
of the two innermost measurements the surface North and South surface
brightness are comparable, suggesting that these two measurements
trace the underlying galaxy velocity field.  Optical velocity results 
are presented in \S~\ref{Sec:ShellKinematics}.

\section{HI observations}
\label{Sec:HIobservations}

Neutral hydrogen VLA observations were made in D array (1 km) in
september 1996.  The VLA data confirmed the presence of HI associated
with the dust lane and southern shell inferred in the WSRT data, and
also reveal the presence of low surface brightness extended HI around
the galaxy.  The HI emission at anomalous velocities in the south
shell region (Fig.~2 of BS96) was not confirmed.  The data looked
complex enough for a proper multi configuration study to allow the gas
to be compared with optical structures on all scales.  Here we present
the combined results of 4 hours of D, 8 hours of C and 16 hours of B
configuration, data taken in September 1996, September 1997 and
February 1997, respectively.  For all observations a total bandwidth
of 6.25 MHz was used, centered at 2870 km/s.  The correlator was used
in 2 IF mode with no online Hanning smoothing.  This produced 63
channels, with 21 \kms\ channel separation.  Useable data were
obtained over the velocity range from 2365 \kms\ to 3521 \kms. 
Standard VLA calibration procedures were used, initially separate data
cubes were made for each of the configurations to inspect the data. 
Then the continuum was subtracted in the UV plane by making a linear
fit through the line free channels.  The UV data were then combined
and image cubes were made with various weighting schemes.  Here we
present the results of the full resolution data, using uniform weight
and robust 1, giving an angular resolution of 7.3x7.2 arcsec and an
rms noise of 0.16 mJy/beam (1~mJy/beam = 11.4~K) and images made of
the C and D array only, with a resolution of 25.2x19.7 arcsec and an
rms noise of 0.23 mJy/beam (1~mJy/beam = 1.21~K).  This translates
into typical column density sensitivities (2 $\sigma$) of $1.4 \times
10^{20}$ cm$^{-2}$ in the BCD array and $2 \times 10^{19}$ cm$^{-2}$ in
the CD array.  Our 6-$\sigma$ HI mass limit over 40 \kms\ is $2 \times
10^7 M_{\odot}$ in the center of the field.

\begin{deluxetable}{lllllll}
\tablecolumns{7}
\small
\tablecaption{HI data for NGC~3656 and detected companions
		\label{Tab:HIPars} }
\tablehead{
\colhead{Name} 	& \colhead{RA (2000)} 	& \colhead{Dec (2000)} &
\colhead{$V_{HI}$} & \colhead{$\Delta(V_{HI})$\tablenotemark{a}} 
& \multicolumn{2}{c}{$M_{HI}$\tablenotemark{b}} \\
\colhead{}	& \colhead{}	& \colhead{}	& 
\colhead{\kms}	& \colhead{\kms} & \colhead{Jy\kms} & \colhead{$10^{9} M_{\odot}$} }
\startdata  
NGC 3656         & 11 23 38.4    & 53 50 30    & 2870 & 425 & 5.40 & 2.0\\
MAPS-NGP O\_130\_0507168        & 11 22 23.6    & 53 51 54    & 3059 & 105 & 0.47 & 0.18\\
MCG +09-19-052   & 11 22 25.2    & 53 41 17    & 2944 & 168 & 3.40 & 1.3\\
MCG +09-19-056   & 11 23 02.5    & 53 41 47    & 2796 & 168 & 4.99 & 1.9\\
MCG +09-19-059   & 11 23 16.0    & 53 47 14    & 2796 & 126 & 0.78 & 0.29\\
UGC 6422         & 11 24 44.7    & 53 44 36    & 2996 & 126 & 0.45 & 0.17\\
\enddata
\tablenotetext{a}{Full velocity range over which HI is detected.}
\tablenotetext{b}{HI mass uncertainties are approximately 10\%.}
\end{deluxetable}

\begin{figure}
\epsscale{1.}
\plotone{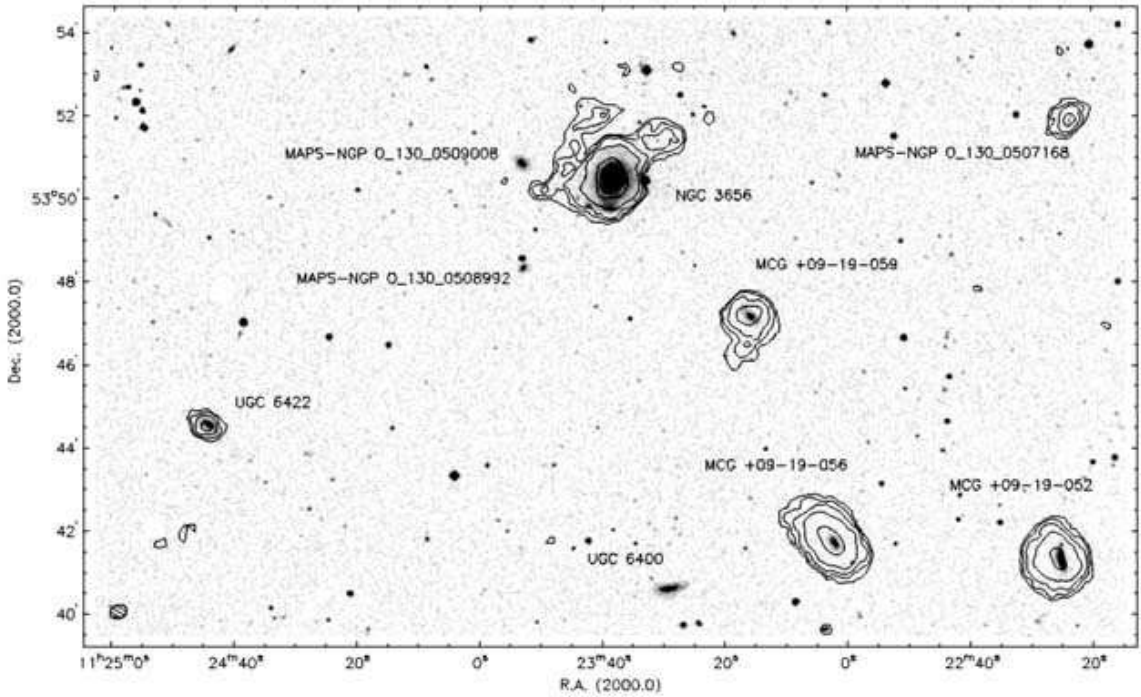}
\caption{ \label{Fig:WideField} Total HI contours from the C and D
array data (25 arcsec resolution) overlaid on a DSS image.  Field size is
$25'\times 15'$.  HI
contours are 0.23, 0.57, 1.1, 2.3, 6.8, 11.4, 16.0 $\times 10^{20}$ 
cm$^{-2}$.}
\end{figure}

\section{Results}
\label{Sec:Results}

HI associated with NGC 3656 is detected over the range of 2600 to 3100
\kms, with an integrated HI flux of $5.4 \pm 0.4$ Jy~\kms, in good
agreement with the single-dish value of $5.3 \pm 1$ Jy~\kms\ obtained
at Jodrell Bank (BS96).  In addition five small spirals and dwarf
irregulars are detected within the velocity range of NGC~3656, all
within $\sim$180 kpc of NGC~3656.  Figure~\ref{Fig:WideField} shows
the total HI distributions for the six detected galaxies over a
Digitized Sky Survey image.  None of the companions had a previously measured redshift.  The
parameters of these galaxies are listed in Table~\ref{Tab:HIPars}. 
Together, the observations indicate that NGC~3656 is in an HI rich
environment.  

MGC~+09-19-059, at 24 kpc to the south-west of NGC~3656,
shows an outer HI extension toward the south. At very
low levels (1-$\sigma$) there is evidence for a bridge between MCG 
+09-19-056 and MCG +09-19-059 at 
2744 \kms, and at even lower levels there may be a connection
between MCG +09-19-059 and NGC~3656 at 2807 km/s.

The HI density distribution of NGC—3656 shows a North-South disk
seen almost edge-on which coincides in the inner part with the dust
band, and extensions on a larger scale surrounding the optical
picture.  At the galaxy center there is a radio continuum source
slightly extended North-South and coinciding with the dust lane.  It
has a peak of 10 mJy/beam and total flux density of $24 \pm 1$
mJy.  This source had been reported by M\"ollenhoff, Hummel \& Bender
(1992).

\begin{figure}
\epsscale{1}
\plotone{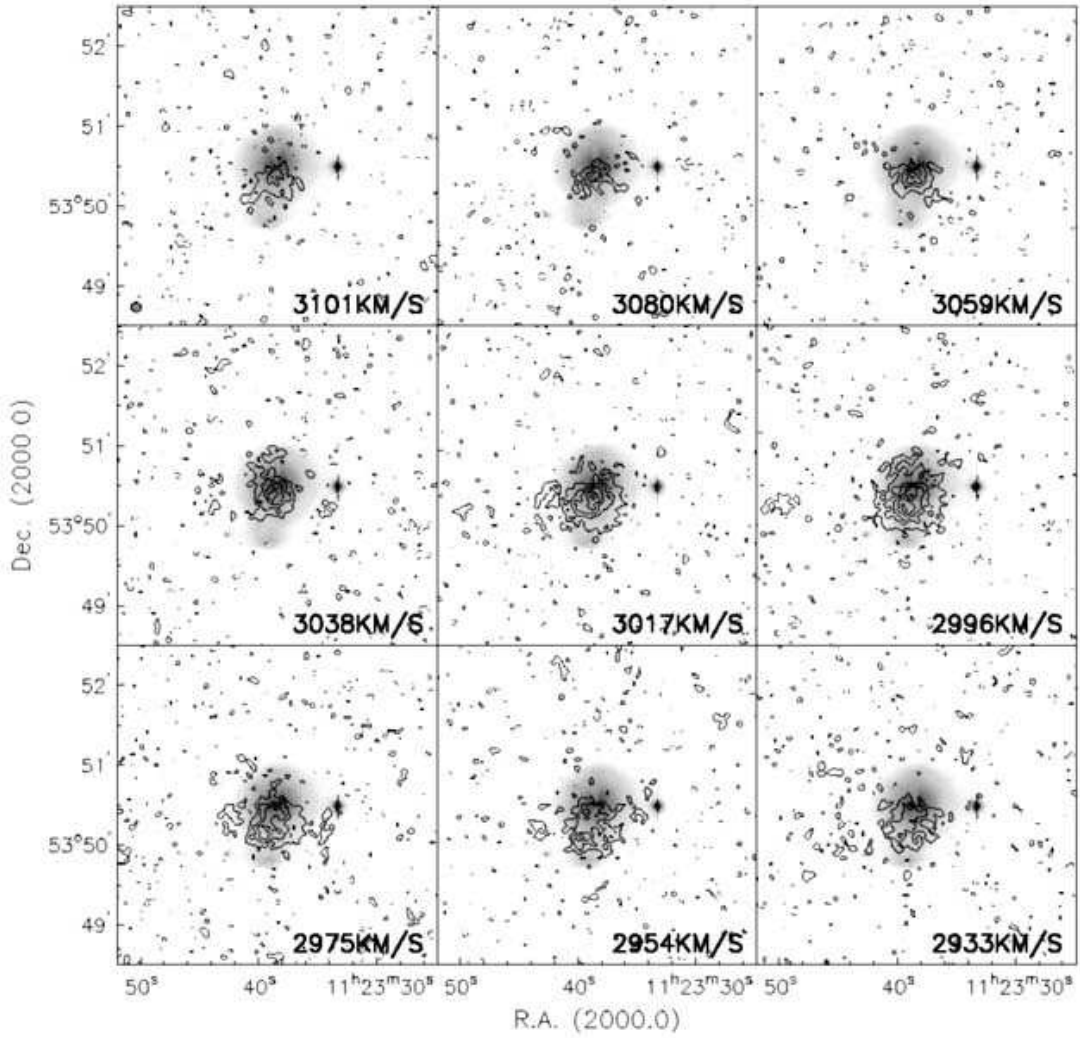}
\caption{ \label{Fig:ChanMaps} Channel maps for the data cube derived 
from BCD array data.  Grayscale is the R band image. Contours 
are -0.64, -0.32 (dashed), 0.32, 0.64,
0.96, 1.128, 1.60 mJy/beam.  Beam is 7.3''$\times$7.2''.  
Noise is 0.16 mJy/beam rms. }
\end{figure}

\begin{figure}
\figurenum{4}
\plotone{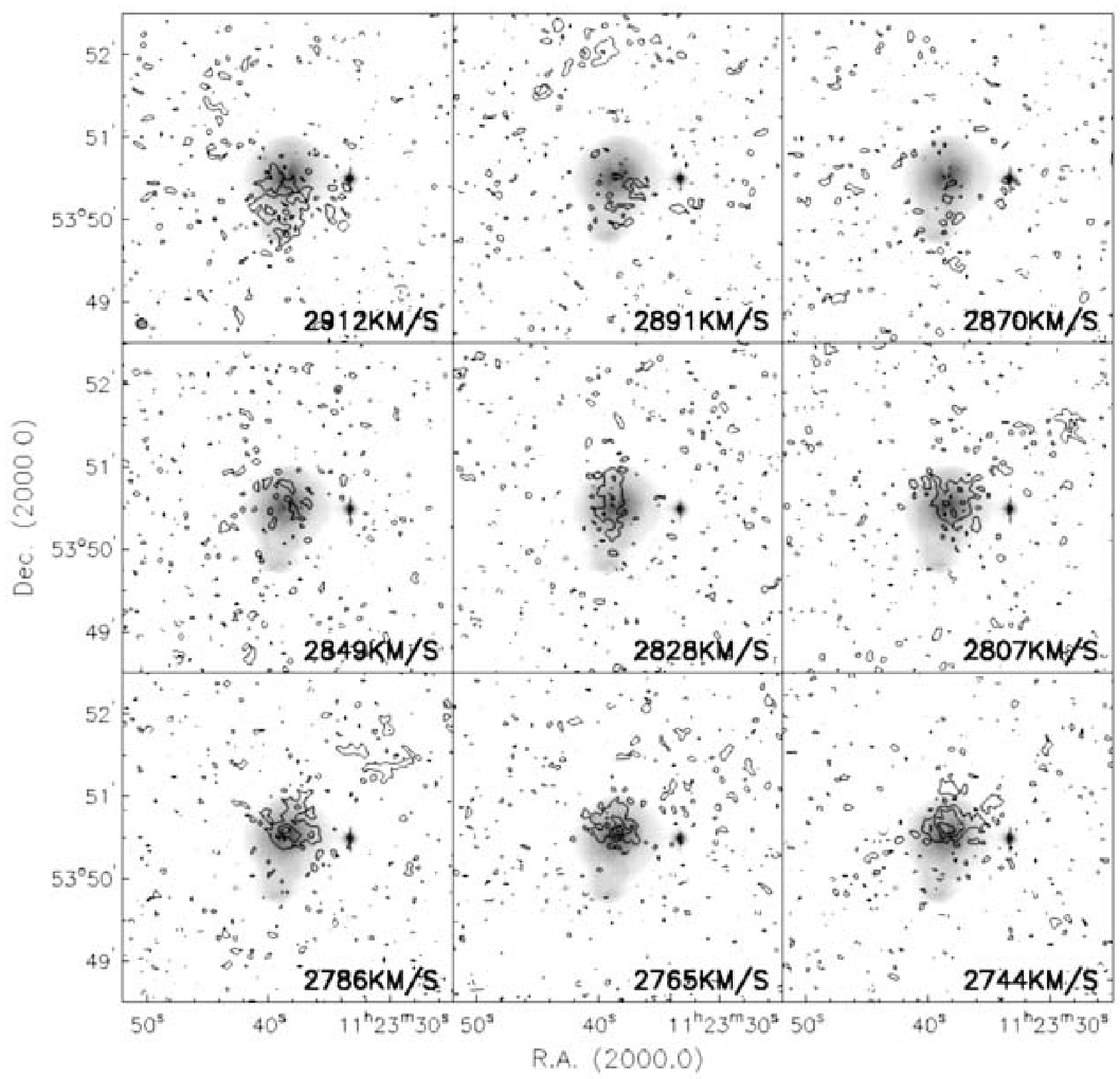}
\caption{ Continued.}
\end{figure}

\begin{figure}
\figurenum{4}
\plotone{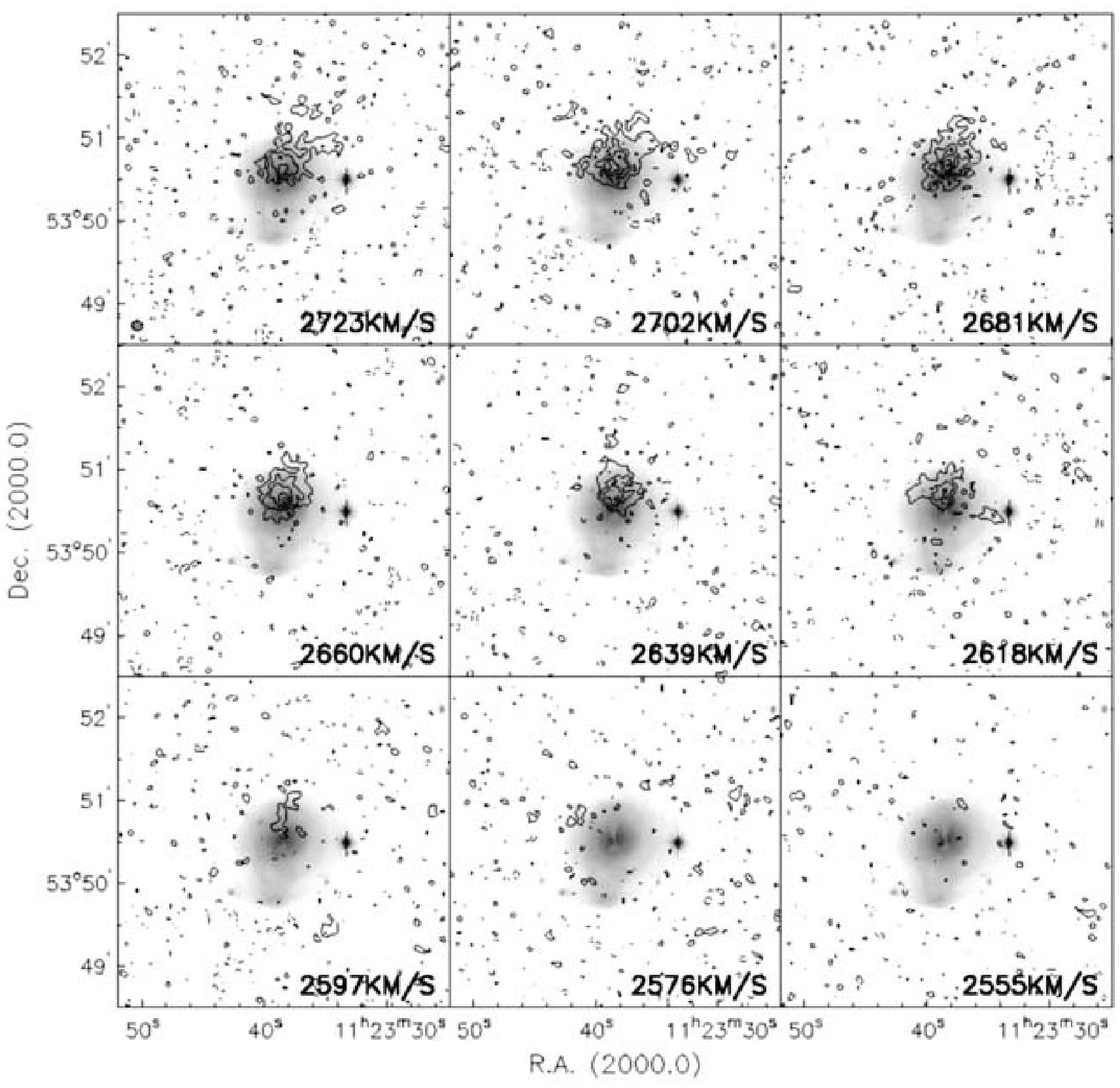}
\caption{ Continued.}
\end{figure}

\begin{figure}
%\epsscale{1.}
\plotone{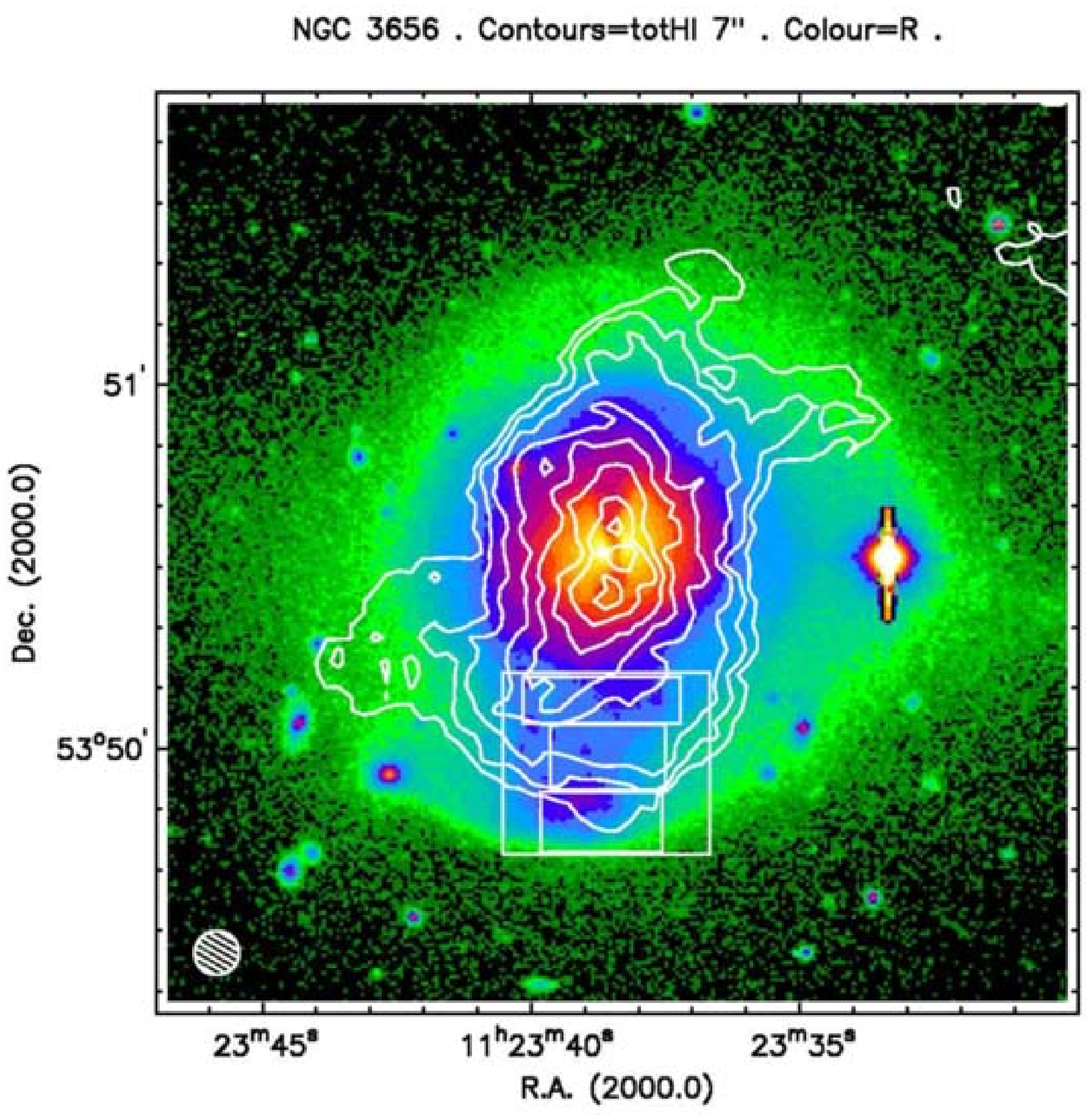}
\caption{ \label{Fig:TotHI7} Contours of total HI flux from BCD
array data.  Beam is 7" FWHM. HI contours are 2.1, 5.2, 10, 21, 31,
42, 52, 62 $\times 10^{20}$ cm$^{-2}$.  {\it Thick box:}
Aperture of the 2D fiber bundle spectrograph.  {\it Thin boxes:}
Synthetic apertures for stellar velocities derived from the 2D optical
fiber spectroscopy.  }
\end{figure}

\subsection{The inner HI disk}
\label{Sec:InnerDisk}

\begin{figure}
\plotone{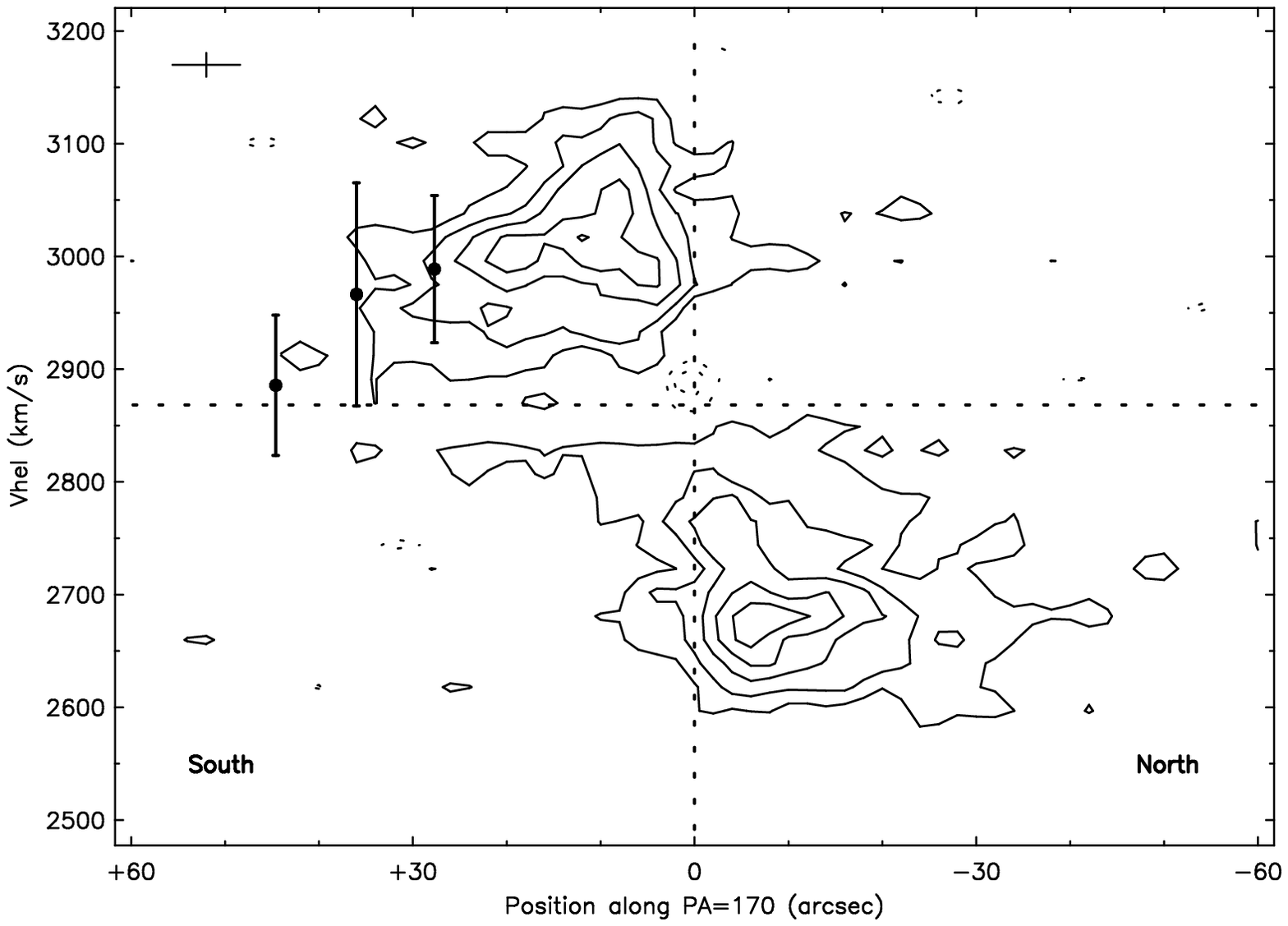}
\caption{ \label{Fig:PosVel7} {\it Contours: } Position-velocity
diagram along position angle 170$^\circ$, centered on the nucleus of
NGC~3656.  The origin is set at the position of the radio continuum
source.  The horizontal dotted line indicates the optically-determined
systemic velocity (2969 km\,s$^{-1}$).  The cross indicates the
spatial resolution (7") and velocity resolution (21 km\,s$^{-1}$). 
Contours are -0.96, -0.64, -0.32 (dotted), 0.32 (2-$\sigma$), 0.64,
0.96, 1.128, 1.60 mJy/beam.  {\it Points: } NaI $\lambda$5890 optical
stellar velocities derived from 2D optical fiber spectroscopy near the
southern shell.  Note the absorption at slightly redshifted velocity.}
\end{figure}

The distribution and kinematics of the gas at the highest angular
resolution of 7 arcsec is displayed in the channel maps of
Figure~\ref{Fig:ChanMaps}.  The high resolution total HI map
(Fig.~\ref{Fig:TotHI7}) shows a concentration of HI running
North-South in the inner parts and well aligned with the dust lane. 
This forms an inner disk seen nearly edge-on.  The central, innermost
contour encircles a local minimum due to absorption against the
central radio continuum source.  Further out the disk seems to become
warped, oriented more North-West/South-East and apparently more
face-on.  The kinematic structure of the edge-on disk is shown in the
position-velocity map at position angle 170\deg\
(Fig.~\ref{Fig:PosVel7}).  This is in the direction from the center to
the southern optical shell.  The pattern of rotation is clear, with
the southern side receding and the northern approaching and a steep
rise in the rotation curve near the center.  The velocity range is
from about 2650 to 3100 \kms, corresponding to 225 \kms\ maximum
rotational velocity, and the mid-point velocity is approximately 2875
\kms, in good agreement with the optically determined stellar systemic
velocity ($2869 \pm 13$ \kms, BS89).  
The warped disk extends to 35 arcsec (7 kpc) from the center.

The disk kinematics is strongly lopsided.  While inside 10 arcsec of
the center, where the maximum rotational velocity of 225 \kms\ is
reached, the rotation pattern is symmetric, beyond 10 arcsec the
rotational velocities remain constant on the northern side whereas
they drop by a factor of 2 ($\sim 100$ \kms) on the southern side.  
This asymmetric velocity
structure indicates large deviations from circular motion -- possibly
gas on eccentric orbits suggesting a young dynamical age for the disk. 
Non-circular motions may contribute to the important velocity
broadening toward systemic velocity visible on both sides of the
galaxy.  On the southern side, low velocity gas (2912 \kms, 35 \kms\
above systemic) is seen at all radii.  There are traces of emission at
apparently counterrotating velocities, approaching in the southern (2800
\kms) and receding in the northern quadrant.  These may partly be
explained by the outer warp geometry.  They may also trace HI in highly elongated orbits associated with the southern shell (see \S~\ref{Sec:ShellKinematics}).

The effect of absorption against the central radio continuum source is
clearly shown by the hole near the systemic velocity in the
position-velocity map (Fig.~\ref{Fig:PosVel7}), as expected in an
edge-on disk.  The absorption is also present at velocities higher
than the systemic velocity (-0.74 mJy at 2891 \kms\ and -0.45 mJy at
2870 \kms.)  This provides evidence for non-circular motions.

\begin{figure}
\plotone{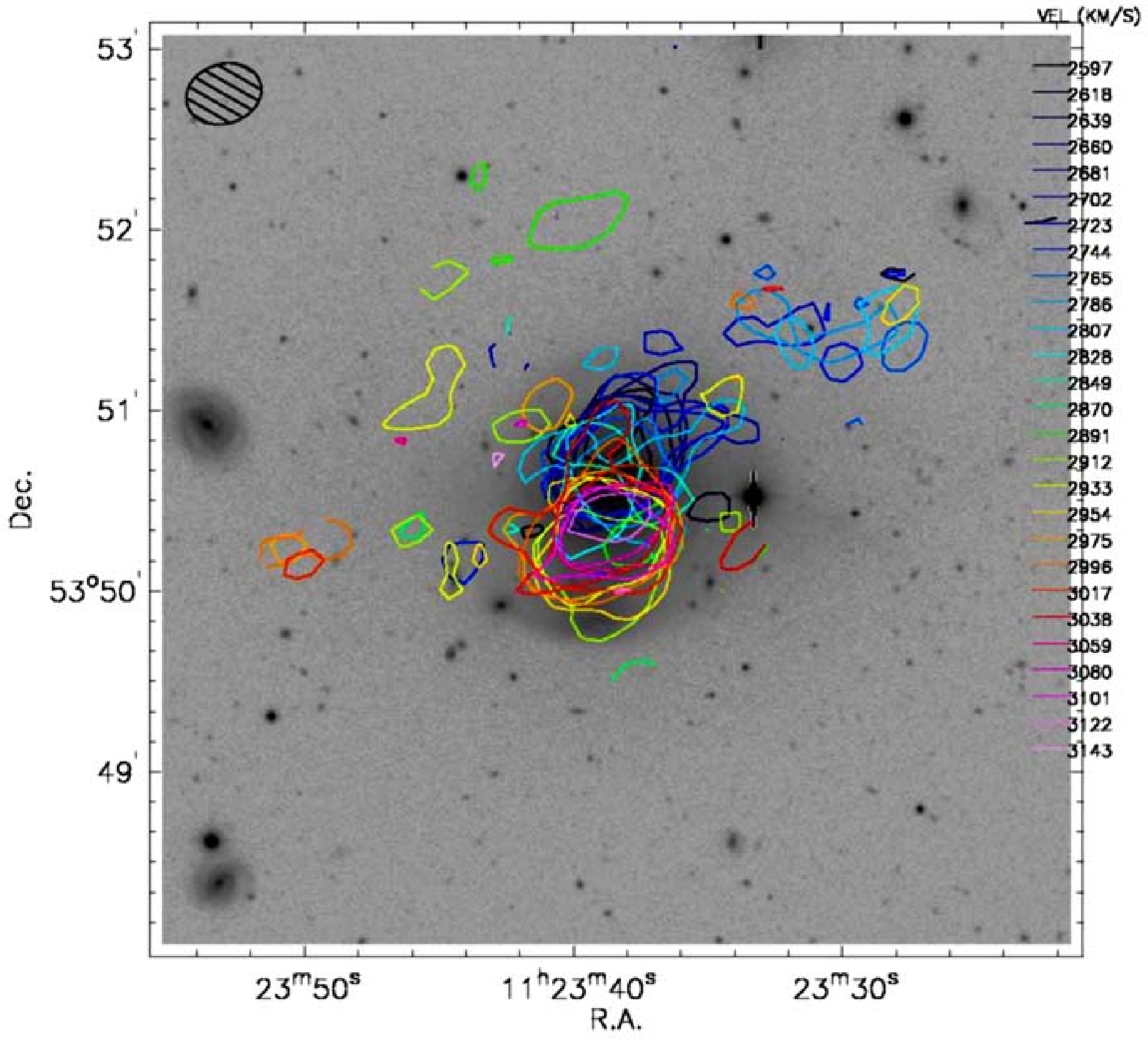}
\caption{ \label{Fig:RgramCD} Contours of flux density 0.64 mJy/beam
in the CD data cube, overlaid on an R-band image of NGC~3656. 
Velocities corresponding to each color are given in the legend.  A
mask was applied to each channel, built from a Gaussian-smoothed total
HI map, in order to highlight the velocity structure of the extended
components.  Beam size is 25.2''$\times$19.7''.  Noise is 0.23
mJy/beam rms.  }
\end{figure}

To investigate the relation of the HI and the shell, we plot in
Figure~\ref{Fig:RgramCD} a Renzogram (Schiminovich, van Gorkom,
\& van der Hulst 2001): one contour (0.64
mJy/beam, 3-$\sigma$) from each channel map, with color-coded
velocity, is plotted on top of a gray-scale $R$-band image.  This
figure may be used to analyze the relation between the rotating system
of HI and the southern shell.  HI is detected at the shell
in the 2912 \kms\ channel only.  The channel maps in
Figure~\ref{Fig:ChanMaps} also show that the HI only reaches the shell
in this channel, at the 4-$\sigma$ level, with traces of HI beyond the
shell in the 2870 \kms\ channel.  The other velocity channels do not
quite reach the shell at the 3-$\sigma$ level.  Most velocity contours
pile up at $\sim$35 arcsec from the center.  Therefore, despite its
proximity, the shell does not appear to lie within the main HI
rotating system, but is slightly offset to the outside of it.  We
find measurable HI on the shell, which does not rotate with the
main HI disk but, rather, displays a narrow range of near-systemic
velocities.

The strong velocity broadening in the disk is apparent in the 
superposition of the velocity contours, as is the warp geometry.  
The color coding shows that the intermediate velocities at the North and South tips of the disk link smoothly with the extended HI emission further out.  

\begin{figure}
\plotone{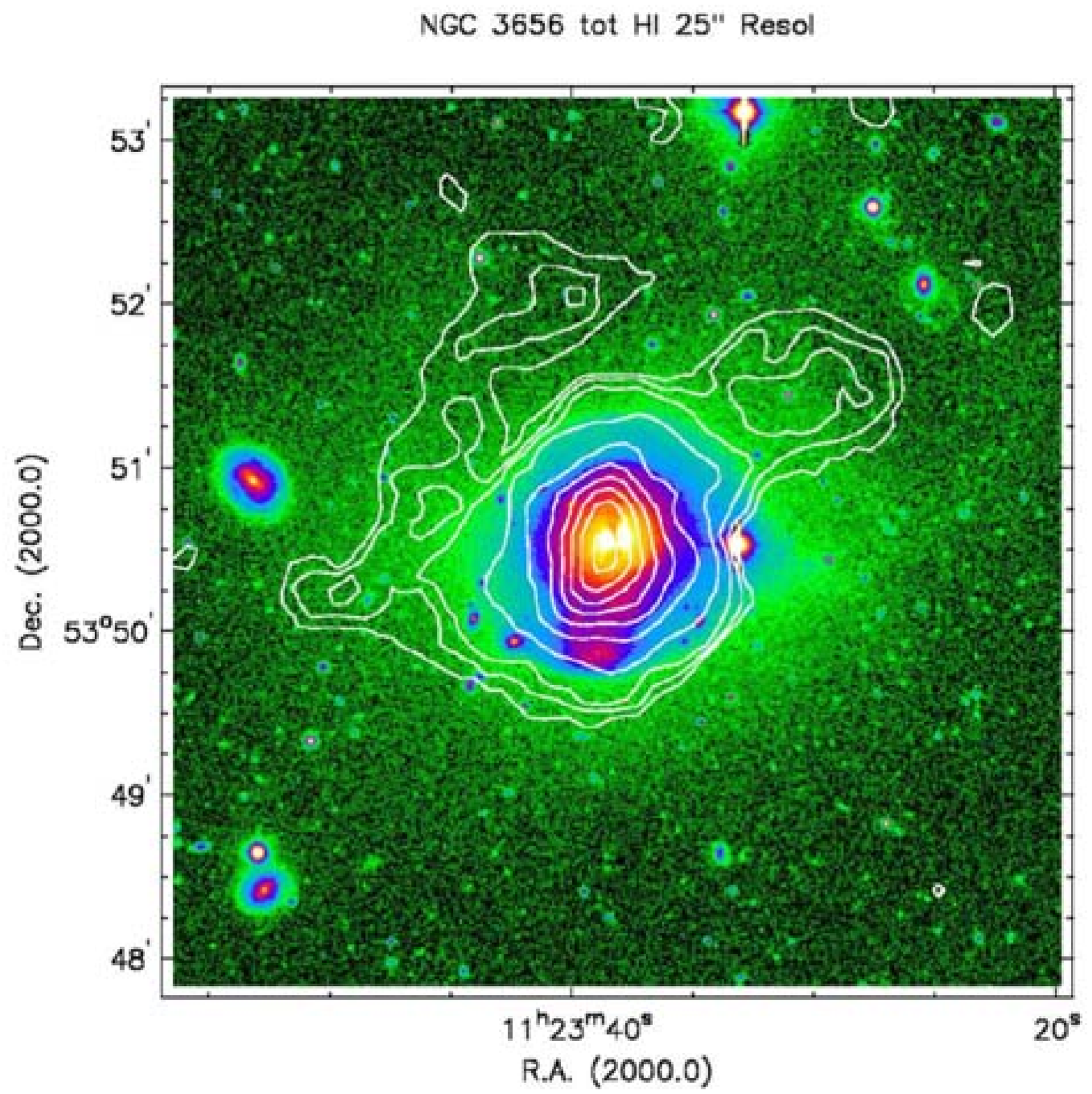}
\caption{ \label{Fig:TotHI25} Contours of total HI flux from the
CD array data.  Beam is 25.2''$\times$19.7''.  HI contours are 0.23,
0.57, 1.1, 2.3, 4.6, 6.8, 9.1, 11.4, 13.7, 16.0 $\times 10^{20}$ 
cm$^{-2}$.  {\it Grayscale: } R-band image of NGC~3656, with
gray scale stretched to show the faintest extended structures.}
\end{figure}

\subsection{Extended emission}
\label{Sec:ExtendedEmission}

The C and D array data reveal HI emission further out than the
$\sim$35 arcsec radius of the main disk.  The most striking
characteristic of the extended emission is its strongly asymmetric
distribution with respect to the optical galaxy.  The total HI 25
arcsec resolution map is shown in Figure~\ref{Fig:TotHI25}, overlaid
on a high-contrast optical image.  While on the southern side the HI
distribution ends rather sharply at the radius of the main disk, two
HI complexes extend to the north.  One starts at the south-eastern
side of the inner HI system and extends like a long arm (130 arcsec =
26 kpc) towards the North.  Its velocity (Fig.~\ref{Fig:ChanMaps})
declines smoothly from about 2933 \kms\ in the East to 2891 \kms\ in
its northern tip.  Another feature starts at the North end of the
inner disk and extends in the North-West direction.  Its velocity
increases from 2723 \kms\ at its base to 2807 \kms\ at its tip.  These
two complexes may be tidal tails, given the asymmetry to the north of
the extended distribution and the smooth velocity profiles along the
features.  An alternative possibility is that they may form part of an
outlying, broken, extended and warped disk or ring.  The observed
velocities may then be understood as due to the more face-on view of
the outer gas as well as to non-circular motions.  These features
connect with the warped gaseous disk in the main body of the galaxy. 
The total HI mass of these two complexes is $1.9 \times 10^8$
M$_\odot$, roughly 9\% of the total HI detected in the system.  The
eastern HI feature is parallel but slightly offset inward with respect
to the eastern optical tail reported by B97, which is shown in
Figure~\ref{Fig:TotHI25}.  The North-West HI feature has no apparent
optical counterpart; the western optical tail curves around the HI
(Fig.~\ref{Fig:TotHI25}).  Such anti-correlation of HI and optical
tails has been found in other merger remnants, eg. NGC~520 (Hibbard
\& van Gorkom 1996) and NGC~2865 (Schiminovich et al.\ 1995).  See 
Hibbard, Vacca, \& Yun (2000) and Mihos (2001) for possible 
explanations of these morphologies.  

\subsection{Shell kinematics}
\label{Sec:ShellKinematics}

Stellar velocities derived from the integral-field optical fiber
spectroscopy (\S~\ref{Sec:2Dspectroscopy}) are shown in
Figure~\ref{Fig:PosVel7} overplotted on the HI position-velocity
diagram.  The two innermost points lie within the rotating system
described in \S~\ref{Sec:InnerDisk}, and their velocities match those
of the HI. These measurements trace the general stellar velocity field
rather than the clump/shell velocity (\S~\ref{Sec:2Dspectroscopy}). 
Thus, stars near the galaxy's minor axis rotate with the HI. The stars
could have been accreted with the HI, or may have formed in place out
of the HI disk.  The pronounced blue colors in this region ($B-R\sim
0.9$, B97) and the traces of H$\alpha$ absorption 
(\S~\ref{Sec:2Dspectroscopy}) 
indicate a young age, and the disk is actively forming
stars further in, out to 3 kpc from the center (BS96).

At the outermost point the mean velocity is $2886 \pm 61$ \kms.  This
point is centered on the bright clump (see
Fig.~\ref{Fig:RbandProfiles}), and thus provides velocity information
closest to the southern shell.  It is interesting that the velocity at
this point lies quite close to the galaxian systemic velocity ($2869
\pm 13$ \kms).  The error bar of the shell velocity, which gives the
rms scatter of the individual fiber velocities, constraints the 
l.o.s. velocity to less than 30\% of the galaxy circular
velocity, which from the overall HI rotation curve we estimate as
$V_{circ} \approx 200$ \kms\ at the galactocentric distance of the
shell.  Even accounting for a comparable velocity component 
perpendicular to the l.o.s., the orbits must be fairly 
elongated.  Trial N-body experiments indicate that, if the shell 
material is at apocenter with tangential velocity below 30 \% of the 
circular velocity, the orbit eccentricity is above 0.8 and the 
pericenter distance is below 0.3$R_{eff}$.  Such radial orbit for the 
clump suggests that the shell results from the piling up of 
stars at the apocenters of their orbits, rather than from the bending 
of a sheet of stars in near-circular orbit seen edge-on at its tangential 
point.  I.e., while the stars cannot be constrained to a strictly radial 
orbit, the shell appears to be more a result of phase-wrapping (Quinn 
1984), than space-wrapping (Dupraz \& Combes 1987, Hernquist \& Quinn 1987).

At the location of the clump in the shell, HI is detected at 2912 \kms, plus
traces further out than the shell at 2870 \kms\
(Figs.~\ref{Fig:ChanMaps}, \ref{Fig:RgramCD}).  These velocities
bracket the optical velocity of the shell.  The strong detection at
2912 \kms\ (above 4-$\sigma$) indicates that the HI at the shell is
dominantly at velocities that match the shell stellar velocity, and
provides kinematic evidence that the HI and the stars are dynamically
associated at the shell.

\section{Discussion}
\label{Sec:Discussion}

The elliptical galaxy NGC~3656 has remarkable optical and HI
properties.  The optical picture is already complex, with a minor axis
dust disk harboring an extended starburst, a North-South blue ring,
two optical tails, a prominent shell, a number
of other faint shells (BS90, B97).  The HI data presented here adds to
the complexity, showing a minor axis warped gaseous disk and two
gaseous complexes or tails extending to the North of the galaxy.  The
HI distribution is asymmetric in two respects: there is a strong
positional asymmetry in the external gas, and a strong kinematic
asymmetry in the HI disk inside the galaxy.  Neutral hydrogen appears
to be dynamically associated to the shell to the South of the
main body.  

Main questions on the HI disk structure are the
pronounced broadening of the position-velocity distribution toward
systemic velocity and the strong velocity lopsidedness
(Fig.~\ref{Fig:PosVel7}).  Modeling will be needed to determine the
relative importance of the warp, the disk internal velocity dispersion
and line-of-sight integration effects on the velocity broadening. 
Modeling may clarify as well if elliptical orbits in the disk or associated with the shell are required to explain the velocity lopsidedness. 

It is likely that the asymmetry in the HI velocity distribution is
also related to accretion from the external complexes, with HI
dominantly falling from the North causing the observed kinematic
asymmetries.  That accretion is taking place may be inferred from the
geometry of the east complex: for this system to wrap around the
galaxy, from its base on the South to its North tip, it needs to be on
the near side to the observer, otherwise its angular momentum would be
opposite to that of the gas disk.  The velocities above systemic then
indicate that the gas in the complex is falling onto the galaxy.  

An order-of-magnitude estimate of the gas accretion rate may be
derived from the masses and velocities of the external complexes.  For
the North-East arm, the rate is $\sim 0.2$ M$_{\odot}$/yr.  At this
rate, both systems would take 5 Gyr to deliver the gas currently
present inside NGC~3656.  Accretion is likely to have been more
efficient in the past, hence the age of the event that originated the
accretion is likely to be much shorter.  If the rate continues as it
is now, NGC~3656 will continue to accrete gas for at least another
$\sim$ 0.5 Gyr, probably longer if the tips of the arms are on less
bound orbits.  

That the external complexes trace tidal tails rather than an external
disk or ring remains open to interpretation, as is the case for the
extended HI distributions in other merger remnants (NGC~520, Hibbard
\& van Gorkom 1996; NGC~5128, Schiminovich et al.\ 1994).  Tails and
inclined rings may have similar kinematic signatures when seen in
projection, despite tails being on fairly elongated orbits.  In
NGC~3656, the strong lopsidedness of the external HI together with the
kinematic peculiarities in the inner disk favor the tail
interpretation.  But it is unclear whether the extended material, or
part of it, describes an extension of the inner warped disk.  
In either interpretation, the long dynamical times in the outer parts make it unclear what the evolution of the gas and stars might be when the merger signatures in the inner parts fade out.  Some of this material may evolve into the type of HI-delineated outer shells seen in eg. NGC~5128 and NGC~2865 (Schiminovich et al. 1994, 1995) -- see in particular the channel maps at  2912 and 2891 \kms.

As the system evolves, the
disk should grow in radius as the high angular momentum material at
high radii settles into organized orbits.  This should happen in about 1
Gyr, roughly the orbital period at the tip of the extended complexes
at 26 kpc.

\subsection{Shell dynamics}
\label{Sec:ShellDynamics}

The first detection of HI near the shells in NGC~5128 (Schiminovich et
al. 1994) brought the question of whether gas and shell stars are
dynamically associated or simply projected along the line of sight (van
Gorkom \& Schiminovich 1997).  Our measurements provide evidence that
the shell stars and the HI in NGC~3656 share a similar line-of-sight
velocity and hence are most likely dynamically associated.

To our knowledge the southern shell in NGC~3656 is the first shell for
which stellar velocity information becomes available.  Our results
show that IFS provides a means for obtaining velocity information of
shells and other extended, low-surface brightness objects by co-adding
over two-dimensional apertures.  Our formal errors are
large, highlighting the difficulty of the measurement, and 
deeper data will be valuable to confirm the results.  We attempted
this measurement on NGC~3656 because the southern shell is
significantly brighter than most shells in other galaxies.  However,
in NGC~3656 the measurement is made difficult by weak line strengths
due to low metallicity and/or dilution by a young stellar continuum
throughout this galaxy (see the spectrophotometric data on NGC~3656 in
Liu \& Kennicutt 1995).  IFS measurements of shell stellar kinematics may
be succesful on massive ellipticals with red shells, even if fainter
than the shell observed here.  

Our measurement highlights the possibility to sort out the
phase-wrapping vs.  space-wrapping dichotomy for shells on the basis
of data.  To date dynamical models for shells have had to rely on
morphological information only to constrain the types of orbits of the shell
material.

It was unexpected to us that the NGC~3656 shell should be on a
phase-wrapping orbit.  There is little evidence for shell interleaving
in this type-2 (Prieur 1990) shell system.  Given the shell's
proximity to the disk of HI, we suspected a low- to
moderate-ellipticity orbit for the shell material, hence a space-wrapping
nature for the shell.  In the similar galaxy NGC~5128, the high
angular momentum of the stellar halo traced by planetary nebulae (Hui
et al.\ 1995) and the high rotation velocity of the HI near the shells
together point to high-angular momentum shells (Schiminovich et al.\ 
1995).  To explain the nearly radial orbit of the shell of
NGC~3656, one possibility is that disruption near pericenter and
dynamical friction remove angular momentum from the collisionless
component of the merging object, so that stars and gas decouple from
each other.  Gas and stars should be expected to follow 
different dynamics during the merger anyway for the gas to form an ordered 
disk.  

Neutral gas in shells has been seen as a problem for the
phase-wrapping mechanism.  SPH modeling (Weil \& Hernquist 1993) shows
that gas-rich shells in radial orbits get easily stripped of their HI
as shells cross paths.  The main shell in NGC~3656 has apparently
managed to keep some of its HI. Stripping efficiency may be reduced if
the gas at the shell, and/or in the disk, is distributed in discrete
clouds; the fate of gas in shells varies significantly depending on
whether the gas in modeled as a continuous medium or as discrete
clouds (Kojima \& Noguchi 1997, Combes and Charmandaris 1999).  The
frequency of gas cloud collisions with the disk may be further reduced
if the disk is geometrically thin.

\subsection{The shell unusual brightness}
\label{Sec:ClumpShell}

The southern shell of NGC~3656 is atypical for its high surface brightness (1.5 mag$_R$\,arcsec$^{-2}$ over the background,  Fig.~\ref{Fig:RbandProfiles}).  Its integrated magnitude, after subtracting a smooth elliptical model of NGC~3656, is about $R=15$, about 5\% of the total $R$-band light of the galaxy.  While its blue colors indicate young stars, the  shell appears also in $K$-band images, hence its high surface brightness cannot entirely be adscribed to young stellar populations.  Taking into account $B-R \sim 1.0$ (B97), the shell has an absolute magnitude $M_B\sim 17$ comparable to that of a typical dwarf galaxy.  This shell is thus produced by a fairly massive component.  The pronounced surface brightness drop beyond the shell indicates that few stars at the shell location presently have orbital energies higher than those of the shell.  Such order is hard to maintain after a pericenter passage near the galaxy center.   These characteristics may probably be explained with a young age for the shell.  The shell material still retains a memory of its initial orbit at present.  Phase wrapping after several orbits might dilute the present shell into a an ensemble of fainter shells with brightness comparable to those typically found in ellipticals.  

With $\sim$5\% of the total galaxy light, presently the shell tidal field might influence the dynamics of the HI disk, perhaps contributing to its kinematic lopsidedness.  N-body modeling should allow to test this hypothesis.

\subsection{Formation scenarios}
\label{Sec:Formation}

NGC~3656 stands out among peculiar ellipticals for the number of
features related to merger and accretion processes occurring
simultaneously in the same object.  What is the origin of so many
peculiarities?

B97 proposes that NGC~3656 formed in a major merger of two spirals,
given the presence of two optical tails.  This merger should have
formed the entire elliptical galaxy, including its system of shells ,
and have dumped the observed neutral gas.  In merger simulations, the
gaseous components of the precursor galaxies form centrally
concentrated disks while 5--50\% of the gas is ejected into extended
tidal components (Barnes \& Hernquist 1996, Barnes 2001), a picture that matches
the HI distribution of NGC~3656.  Spiral-spiral mergers produce polar
gaseous rings under certain orbital configurations (Bekki 1998).

Shell formation in major mergers was modeled with N body techniques by
Hernquist \& Spergel (1996).  In their scenario, shells form out of
returning tidal material.  In NGC~3656, that the brightest shell lies
at the base of one of the tidal tails and near the HI disk provides
support for a connection of the shell to the gas accretion.  That such
subsystem would have a sharp edge may be caused by the narrow ranges
of orbital energy and angular momentum of the HI returning to the
galaxy along the tidal tails.  The blue colors in the shell also
support the connection between the shell and the gas accretion.  The
lack of H$\alpha$ emission, indicating the absence of early B-type
stars, places a lower limit to the age of the shell stars at $\sim
10^7$ yr (eg.  Leitherer et al.\ 1999), while the presence of
H$\alpha$ in absorption places a less-well defined upper limit at
$\sim$1.5 Gyr.  The galaxy's crossing time at the shell radius is 0.6
Gyr.  Hence, the spectral signatures at the shell indicate ages that
are typical of advanced merger remnants, ie a few remnant's crossing
times.  Note that these ages are consistent with some of the stars
being born in place after the accretion event, perhaps by compression
of the gas clouds in their passage through the inner galaxy.  The role
of star formation in building shell systems has been largely
unexplored, and could be relevant to explaining the puzzling color
properties of shells (Prieur 1990).

Overall, the major-merger hypothesis provides a consistent picture for
the formation of the merger signatures of NGC~3656 in a single
process.  Other interpretations are also plausible, in which different
processes account for the various peculiarities.  The blue colors of
the shell ($B-R \sim 1.0$, B97) correspond to those of a late-type,
low-metallicity galaxy, and thus could trace the ingestion of a
satellite by NGC~3656.  Several gas-rich galaxies exist in the
vicinity of the galaxy, with sufficient gas to provide the $\sim10^9$
M$_\odot$ of HI present in the NGC~3656, while the accretion of other
satellites could have produced the optical tidal tails and the shells. 
The presence of several gas rich companions to HI-rich ellipticals is
common.  For the NGC~3656 group to be bound, using the mass estimators
of Heisler et al.\ (1985), we require from $3\times 10^{12} M_{\odot}$
to $7.4\times 10^{12} M_{\odot}$, yielding mass-to-light ratios
(M/L)$_{B}$ from 125 to 300 in Solar units (adding the light of all
galaxies within 200 kpc and using our own photometry of NGC~3656, the
dominant contributor).  The halo mass is typical for L$_{\star}$
galaxies (Zaritsky et al.\ 1997), suggesting that the NGC~3656 group may
indeed be bound.  NGC~3656 may represent a case of a multiple
accretion onto an existing elliptical.  Such process may continue in
the future with the accretion of the five galaxies that lie within 180
kpc of NGC~3656, representing the collapse of a loose group leading to
the formation of a massive field elliptical.

NGC~3656 shares striking similarities with the nearby elliptical
NGC~5128, the parent galaxy to the radio source Cen~A. B97 reviews the
optical properties of these two galaxies, which have similar
luminosities, effective radii, velocity dispersions, minor axis dust
disks and systems of shells.  The present paper shows that the
similarities between these two galaxies extend to their distributions
of HI, which both show a warped minor-axis rotating disk that extends
outward to link with HI filaments aligned with the optical shells
(Schiminovich et al.\ 1994, see also Sparke 1996).  Evidence for star
formation in the shells is given in NGC~3656 by the spectral
(\S~\ref{Sec:2Dspectroscopy}) and color (B97) signatures, and in
NGC~5128 by the detection of CO in the shells (Charmandaris et al.\
2000).  These strong similarities might indicate that the type of
merger and the nature of the progenitors is similar for these two
galaxies, providing support for the hypothesis of a spiral-spiral
merger for the formation of NGC~5128 (Schiminovich et al.\ 1994, 
Schweizer 1998). 

NGC~3656 may be at an earlier stage in the merger evolution, as judged
by the higher asymmetry of its optical image given by the prominent southern
shell.  Hence we place NGC~3656 in an earlier stage than
NGC~5128 in the merger age sequence of shell galaxies, and we note that
the faint optical tails, HI complexes and shell colors in NGC~3656
make a useful contribution toward filling the ``King gap'' (Toomre
1977) between merger remnants and normal ellipticals.
 
\acknowledgements{We thank Francoise Combes for her fast and constructive refereeing, and M. Vogelaar for his assistance with the use
of the GIPSY image processing system.  This work has been supported in
part by NSF grant AST-97-17177 to Columbia University.  We acknowledge the use of NED and of the LEDA database, http://leda.univ-lyon1.fr/.  }

\end{document}